\documentclass[aps,pre,twocolumn,groupedaddress,superscriptaddress]{revtex4}
\usepackage{graphicx,amsmath,txfonts}

\begin{document}

\title{Ranking influential spreaders is an ill-defined problem}
\author{Jain Gu}
\affiliation{Department of Energy Science, Sungkyunkwan University, 440--746 Suwon, Republic of Korea}
\author{Sungmin Lee}
\affiliation{Department of Energy Science, Sungkyunkwan University, 440--746 Suwon, Republic of Korea}
\author{Jari Saram\"aki}
\affiliation{Department of Computer Science, FI-00076 AALTO, Aalto University, Finland}
\author{Petter Holme}
\affiliation{Department of Energy Science, Sungkyunkwan University, 440--746 Suwon, Republic of Korea}

\begin{abstract}
Finding influential spreaders of information and disease in networks is an important theoretical problem, and one of considerable recent interest. It has been almost exclusively formulated as a node-ranking problem---methods for identifying influential spreaders rank nodes according to how influential they are. In this work, we show that the ranking approach does not necessarily work: the set of most influential nodes depends on the number of nodes in the set. Therefore, the set of $n$ most important nodes to vaccinate does not need to have any node in common with the set of $n+1$ most important  nodes. We propose a method for quantifying the extent and impact of this phenomenon, and show that it is common in both empirical and model networks.
\end{abstract}

\maketitle

\section{Introduction}
Rumors, opinions, ideas and infectious disease all spread on networks. To maximize the impact of word-of-mouth marketing or to contain infectious disease outbreaks, it is essential to identify important spreaders---i.e.\ people that acquire the spreading agent easily and are expected to pass it on to many others. The importance of an individual depends on many factors---the details of disease transmission (we focus on infectious disease spreading from now on), the in-host disease dynamics, and the network structure of the contact patterns, among others. This is the motivation behind the emerging field of network epidemiology~\cite{keeling_rev}.

\begin{figure}
  \includegraphics[width=0.8\columnwidth]{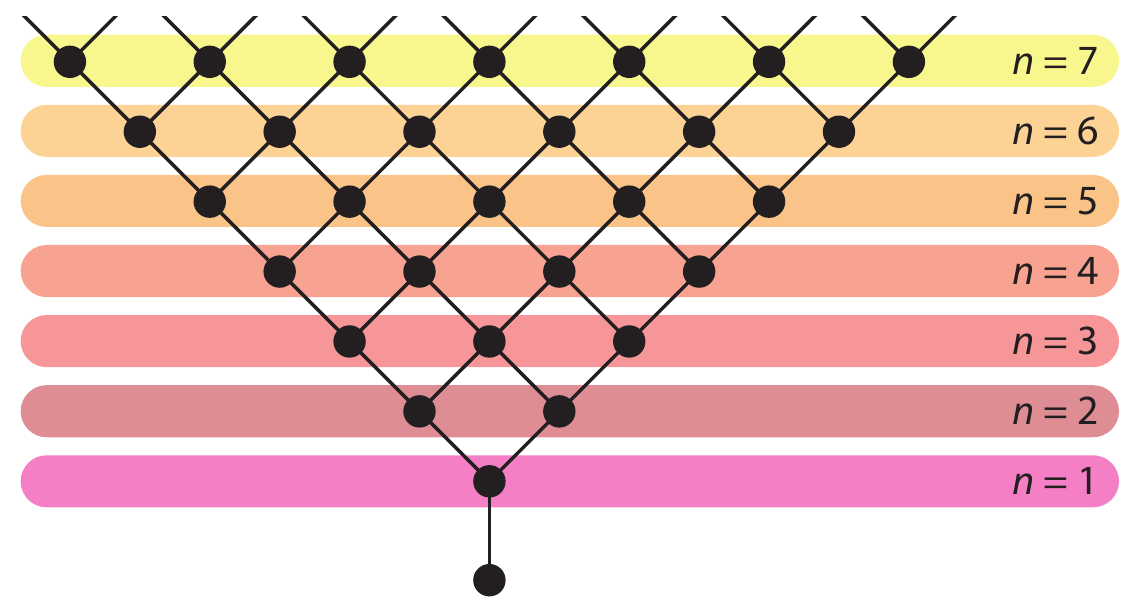}
  \caption{An example of an infinite network where one cannot rank influential spreaders with respect to the reduction of the size of the largest connected component. Let $D_n$ be the set of $n$ nodes that maximizes the number of elements in $\Lambda(X)$---the set of nodes no longer in the largest connected component after the set $X$ is deleted. In this example, $D_n \cap D_{n'} = \emptyset$ for any $n'\neq n$.
}
\label{fig:ex}
\end{figure}

Many studies---e.g.\ Refs.~\cite{nv,holme_vacci,holme_tempo_vacci,core,ci,liu,chen,bramson}---have devised methods to rank nodes according to their importance. As there are many ways to vary the underlying assumptions about the structure of contact patterns, the objective function (i.e.\ how to measure the severity of a disease outbreak), the disease dynamics, and the information available to exploit these structures, these methods are becoming a very rich and diverse theory~\cite{lv}. Typically, it is implicitly assumed that for vaccination or quarantine, the nodes of a network can be ranked with respect to the objective function: if $n$ nodes are to be vaccinated or quarantined, the optimal choice is to always take the top $n$ nodes of the ranking. In this paper, we show that this is not the case. In other words, one cannot, strictly speaking, rank influential spreaders. See Fig.~\ref{fig:ex} for an example where the $n$ most influential nodes (in the sense that deleting them would reduce the largest connected component as much as possible) are not among the optimal $n'$ to delete for any $n'\neq n$. In this work, we show that similar situations arise in many networks, besides this extreme and contrived example case. To this end, we explore how common deviations are from situations where the optimal sets $D_n$ of $n$ nodes to delete are fully nested (i.e.\ $|D_{n+1}\cap D_n|=n$). We first derive a quantity (\textit{ill-definedness}) that measures the extent of such deviations, then show that ill-definedness is common in simple model networks, and finally make the point that the issue persists even in real-world networks. We also address the issue of degeneracy of the optimal sets.

\section{Preliminaries}

We start by introducing some notation. We define a \textit{well-defined} scenario to be one where one can rank vertices according to influence and where the $n$ most influential nodes are always the first $n$ nodes of that ranking. Given a measure of the severity of a disease outbreak (such as the number of nodes that eventually get the disease), let $D_n$ be an optimal set of $n$ nodes to delete with respect to the severity measure, and let 
\begin{equation}
Y_n=\left\{D_n^i\right\}_{i=1}^\nu
\end{equation}
be the set of all optimal sets of $n$ nodes, with $\left\vert Y_n\right\vert= \nu$. An optimal set comprises the $n$ nodes that, if deleted (vaccinated), reduce the severity of the disease as much as possible. The degeneracy $\nu$ of the optimal set is introduced for situations where there is more than one optimal set; this degeneracy depends on $n$. Now let
\begin{equation}
a_n(i)=\min_j\left|D_{n+1}^j\setminus D_n^i\right|-1
\end{equation}
and let $\alpha(n)$ be the average value of $a_n(i)$ over all optimal sets in $Y_n$. If the influence ranking problem is perfectly well defined for a network, then $a_n(i) = 0$ $\forall$ $i,n$ and subsequently $\alpha(n)=0$ $\forall$ $n$, that is to say, the optimal sets of $n+1$ nodes totally include the optimal sets of $n$ nodes for any value of $n$. Conversely, the less well defined the ranking, the larger the value of $\alpha$. We call $\alpha$ the \textit{ill-definedness} of a network. One can interpret $\alpha$ as the average number of nodes that deviate from the well-defined case.

\section{Results}

\subsection{Model networks}

To get an understanding for how common ill-defined rankings are and how the ill-definedness depends on the network, we first study $\alpha(n)$ on small model networks that allow exhaustive treatment. As the severity measure, we use the size $S$ of the largest connected component. The network models we use are $N=L\times L$ square grids and Erd\H{o}s-R\'enyi random graphs~\cite{newman:book}. For the square grids, we use open boundary conditions---node $(x,y)$ of the grid is connected to $(x+1,y)$ unless $x=L$ (and, similarly, $(x,y)$ and $(x,y+1)$ are connected for $0\leq y<L$). For the random graphs, we start with $N$ isolated nodes, go through all pairs of nodes and add links with probability $p$. To calculate $\alpha(n)$, we perform an exhaustive search for optimal sets $D_n$ for the entire range $n\in [1,N]$. As this is computationally very heavy, we restrict ourselves to very small networks that nevertheless clearly illustrate the issue. In the first analysis, we use $N=9$ (i.e.\ $L=3$ square grids).

\begin{figure*}
  \includegraphics[width=0.8\textwidth]{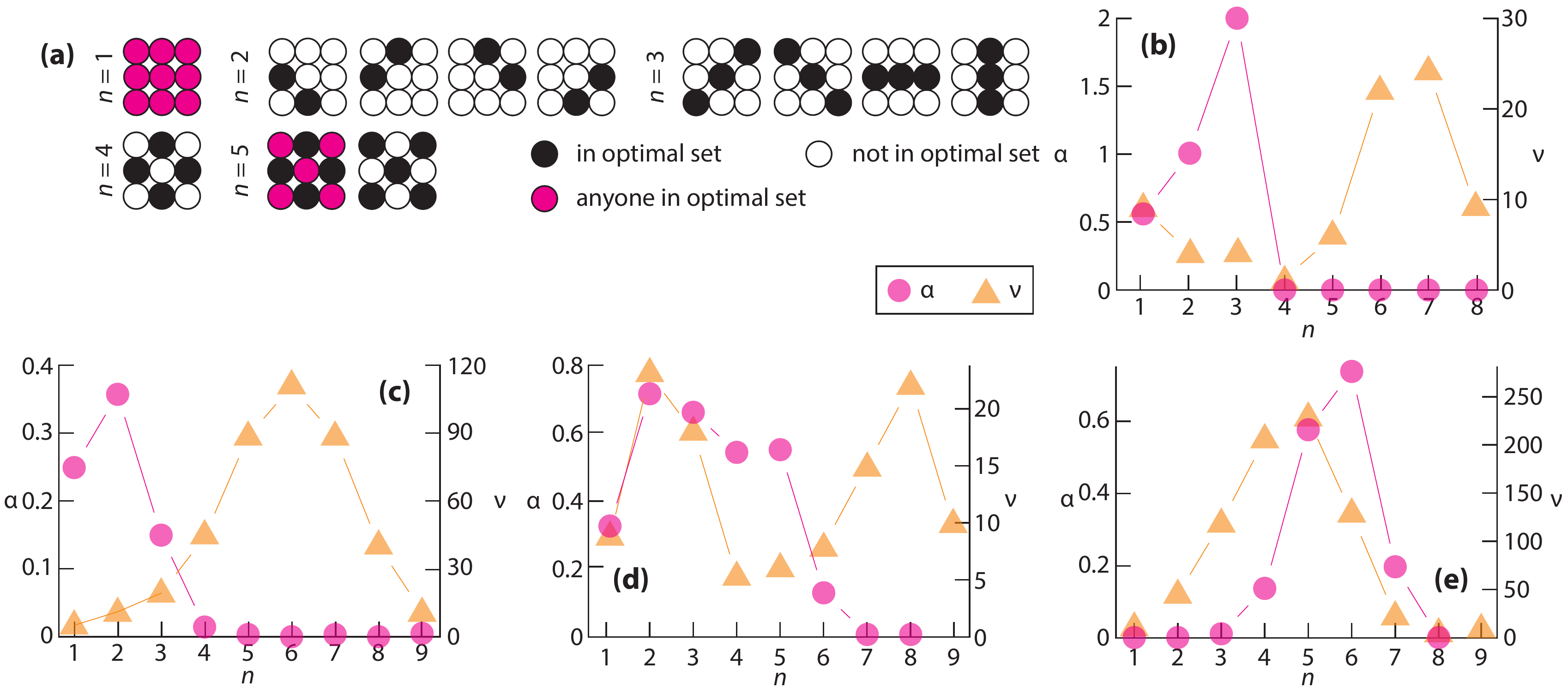}
  \caption{The ill-definedness $\alpha$ and degeneracy $\nu$ as functions of $n$ on a square lattice (b) and on random networks with $p=0.1$ (c), $0.5$ (d), $0.9 $(e). The number of nodes, $N$, is $9$ for the square grid and $10$ for the random networks.}
\label{fig:alpha}
\end{figure*}

For the square grid, the ill-definedness $\alpha$ has its maximum at $n=3$, dropping down to $\alpha=0$ as $n$ reaches $4$ (see Fig.~\ref{fig:alpha}b). Node importance rankings are thus ill-defined for $n<4$. The $3\times 3$ square grid is simple and symmetric enough to understand in some detail (Fig.~\ref{fig:alpha}a). In this case, $D_3$ consists of four sets of nodes---the two diagonals $\{(1,1),(2,2),(3,3)\}$ and  $\{(3,1),(2,2),(1,3)\}$, and the middle row $\{(2,1),(2,2),(2,3)\}$ and column $\{(1,2),(2,2),(3,2)\}$. The degeneracy is thus $\nu=4$. Deleting any of these sets reduces $S$ from nine to three. However, for $n=4$ there is only one optimal set consisting of the center nodes of each side---$(1,2)$, $(2,1)$, $(2,3)$ and $(3,2)$---and so $\nu=1$. When these nodes are deleted, all other nodes are isolated. Thus for $n>4$, deleting these four nodes and any other node in addition would also disconnect the entire network. This also means that, for $n>4$, any $D_n^i$ is one node added to $D_{n-1}^i$. Therefore, for $n>4$, $a_n(i)=0$ for any $i$ and subsequently $\alpha(n)=0$.

The ill-definedness values for random networks ($N=10$, averaged over $1,000$ networks) are shown in Fig.~\ref{fig:alpha}(c), (d) and (e). For the lowest network density ($p=0.1$), $\alpha(n)$ is seen to follow a similar peaked shape for low $n$ as for the square grid, even though its maximum value is smaller. As the networks get denser, the peak shifts towards larger values of $n$. This reflects the fact that it takes more node deletions to disconnect a denser network. For the dense networks of Fig.~\ref{fig:alpha}(e) with $p=0.9$, there are no sets of one or two nodes whose deletion would fragment the network, and thus any set of one or two vertices is optimal and  $\alpha(n)=0$ for $n<3$. The degeneracy has a peculiar dependence on the density of the networks. $\nu(n)$ has one peak for the sparsest (Fig.~\ref{fig:alpha}(c)) and densest networks (Fig.~\ref{fig:alpha}(e)) and two peaks for the networks of intermediate density (Fig.~\ref{fig:alpha}(d)). To understand this, note that $\nu(n)=\binom Nn$ if all nodes are equivalent, which is true for the limiting cases of a network without links and a fully-connected network. Let $n'$ be the value of $n$ above which the network is typically completely fragmented. For $n>n'$, the $\nu(n)$ curve would be peaked for the same reason why the $\nu(n)$ of a network of isolated nodes is peaked (indeed in the same way as discussed for the square grid)---it represents the number of sets to fragment the network plus the number of ways to delete the isolates. The intermediate minimum tells us that when $n$ becomes just so large that fragmenting the network completely is possible, then the sets of vertices to delete to achieve this are few. There are two effects that explain the first peak, \textit{i.e.}~why for small $n$ the degeneracy $\nu$ grows with $n$. First, the number of combinations of $n$ elements out of $N$ increases with $n$. Thus, for homogeneous networks (like the square grid, and to some extent also the random networks) this leads to an increase of $\nu$. Second, for heterogeneous networks---where the degree distribution is very skewed---the top influencer would be very obvious. One would need to continue to higher $n$ before any degeneracy would be at all likely.

\begin{figure}
  \includegraphics[width=0.7\columnwidth]{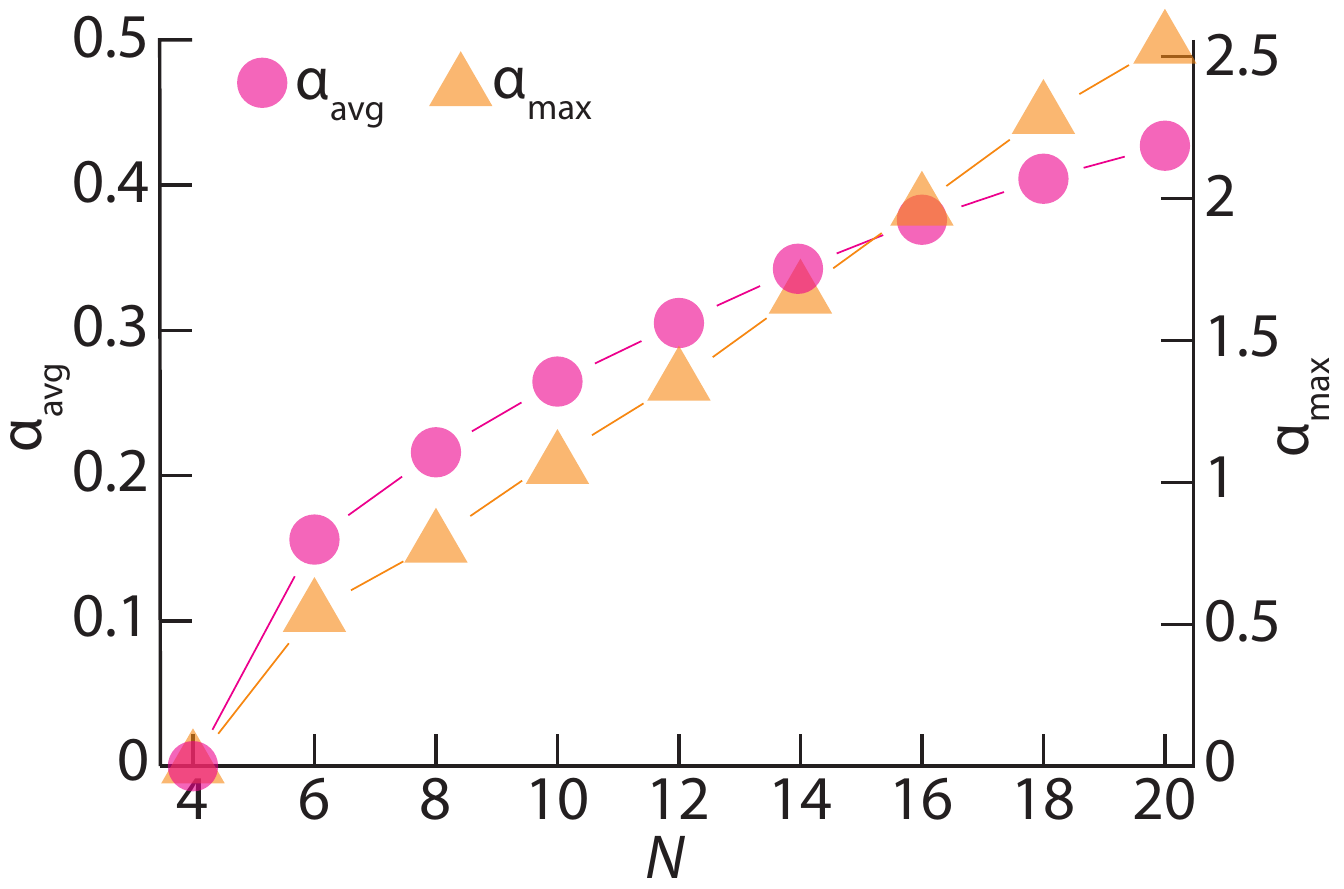}
  \caption{The average and maximum $\alpha$ over all $n$ as a function of network size $N$. The underlying networks are ER model networks with $p=0.5$. The curves are averaged over $>10^3$ networks. Error bars would be smaller than the symbol size and are not shown. 
}
\label{fig:fss}
\end{figure}

Ill-definedness is not limited to the small networks discussed above---rather, it seems to persist as the network size increases. In Fig.~\ref{fig:fss}, we show how $\alpha_{\rm avg}=(1/n)\sum_n \alpha_n$ and $\alpha_{\rm max}=\max_n \alpha_n$ depend on the network size $N$ for ER networks with $p=0.5$. Both of these quantities are increasing. Finally, we have investigated other model networks of varying size, all showing single-peaked $\alpha$ curves and $\nu$ curves with one or two peaks.
 
\subsection{Empirical networks} 

\begin{figure*}
  \includegraphics[width=0.8\textwidth]{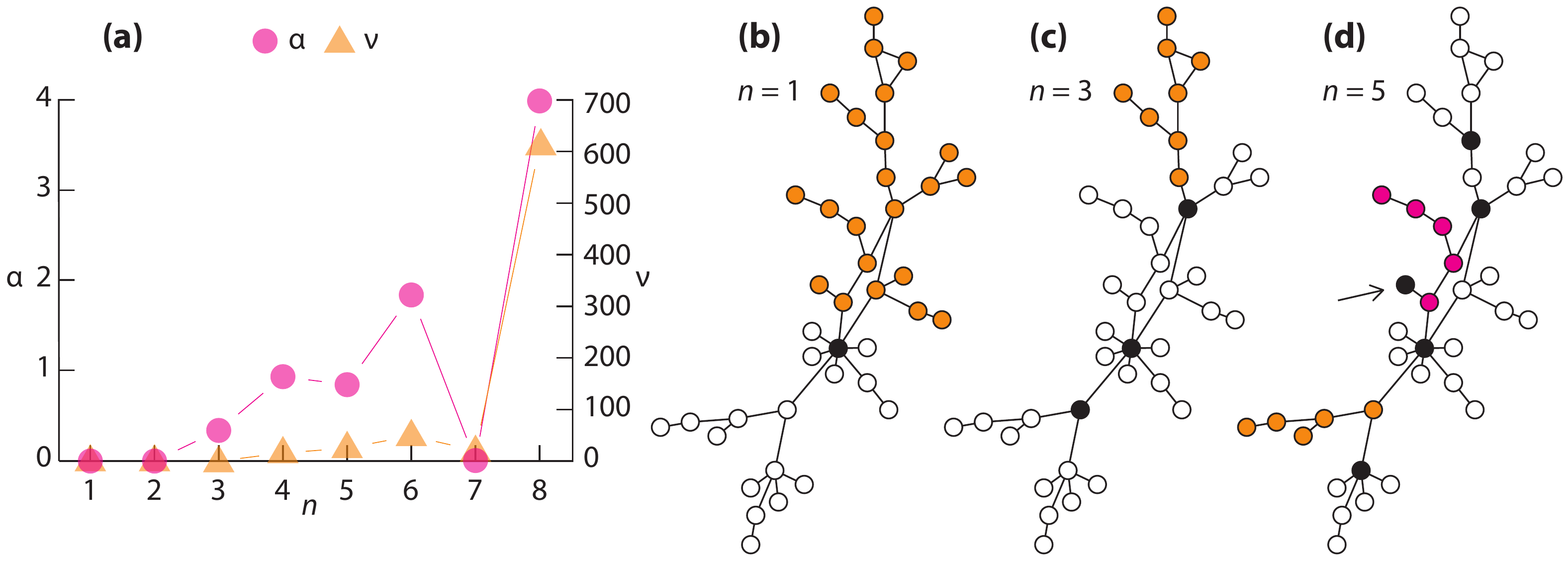}
  \caption{Panel (a) shows the ill-definedness $\alpha$ and degeneracy $\nu$ as functions of $n$ on an empirical network of sexual contacts. Panels (b), (c) and (d) show optimal set of nodes (black) to delete for $n=1$, $3$ and $5$ respectively. The shaded areas of nodes are members of a largest connected component. The highlighted node in (d) is an example of an optional node in the optimal set. It could be replaced by any other node of the largest connected component it is connected to.}
\label{fig:hiv}
\end{figure*}

The real networks that diseases spread over are believed to have much more complex structure---heterogeneous degree distribution, community structure, abundant triangles, etc.~\cite{coregroup,school,hospital,prostitution} We have also investigated some empirical contact networks from the network epidemiology literature. Due to computational constraints we have not been able to scan the full range of $n$, but rather study $\alpha(n)$ for the very lowest values of $n$ only. In Fig.~\ref{fig:hiv}, we show results for a network of sexual contacts from the article first arguing that HIV is a sexually transmitted infection~\cite{hiv}. It is a small network of only $N=40$, still being more heterogeneous than the above-studied random networks. We see a general growing trend of $\alpha$, with a sudden dip to zero at $n=7$. Some specific $D_n^i$ sets are shown in panels Fig.~\ref{fig:hiv}(b), (c) and (d) (for $n=1$, $3$ and $5$ respectively). In Fig.~\ref{fig:hiv}(d) we can see a typical reason for large degeneracy $\nu$. The node highlighted by an arrow could be replaced by any other node in the (shaded) largest component that it is attached to. The actual values of $\alpha$ that we observe in Fig.~\ref{fig:hiv} are larger than for the model networks of Fig.~\ref{fig:alpha}, even though we have only investigated $\alpha$ for very small $n$---the largest $\alpha$ is likely larger. Our preliminary results suggest that in general the average and maximum $\alpha(n)$ values increases with network size $N$. However, computational reasons prevent making a comprehensive study of $\alpha$'s $N$ dependence.

\subsection{Outbreak size in SIR model}

\begin{figure}
  \includegraphics[width=0.5\columnwidth]{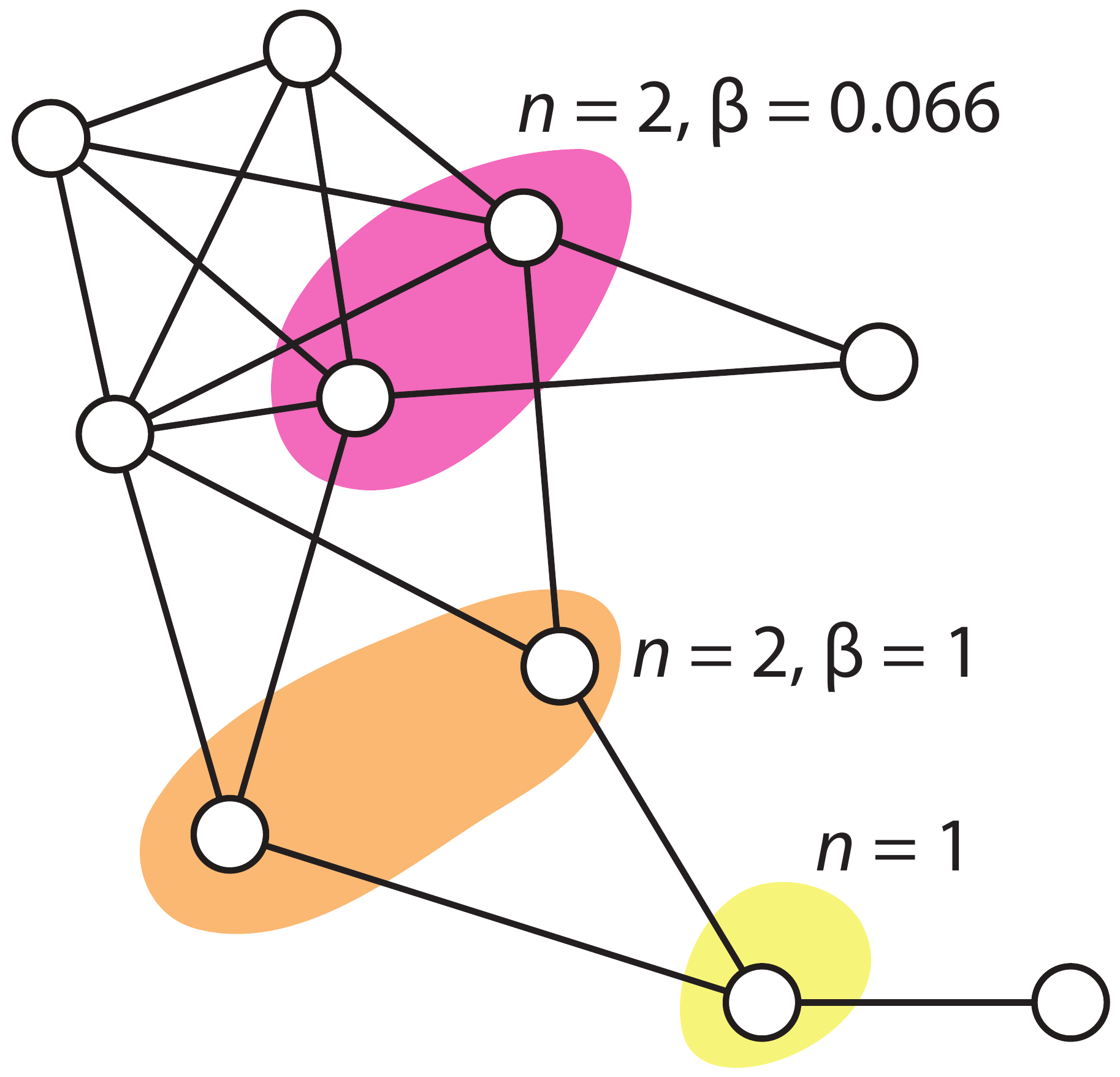}
  \caption{Results for SIR disease spreading on a small example network (drawn from the Erd\"{o}s-R\'{e}nyi random graph ensemble with $N=10$ and $p=0.5$). The optimal sets to vaccinate are shown for two values of $\beta$---$1$ and $0.05$.}
\label{fig:sir}
\end{figure}

In our final numerical study, we investigate a more realistic severity measure than the size of the largest connected component, namely the expected outbreak size $\Omega$ in disease simulations on the network. Since the disease simulations make the analysis yet more computationally demanding, we will only show an example network where the ranking according to which to vaccinate nodes is ill-defined. For disease simulations, we use the SIR (susceptible-infected-recovered) model. This is a standard model of diseases that give the infected person immunity upon recovery~\cite{hethcote}. It starts from a situation where all nodes are susceptible to the disease except one randomly chosen seed node, who is infected. Nodes have a chance $\lambda$ to recover at any unit of time (we set $\lambda=1$). When an infected node is a neighbor of a susceptible node, the susceptible can become infected with a probability $\beta$. We scanned several Erd\H{o}s-R\'{e}nyi random graphs (as above) with $N=10$ and $p=0.5$. We ran $10^6$ outbreaks for every set of $n$ nodes to delete and a range of $\beta$ values. One challenge to analyze this severity measure is that one cannot identify degenerate optimal sets (i.e.\ when $\nu>1$). In the simulations, $\Omega$ can differ for these sets because of stochastic fluctuations, even though they should in theory be equal. Instead of actually measuring $\alpha$, we will just show that $\alpha$ can be larger than zero (i.e.\ the ranking problem is ill-defined). This is illustrated in Fig.~\ref{fig:sir}, where we show an example of a graph where the optimal sets for $n=1$ and $n=2$ are not overlapping (we can say with $>99\%$ confidence that these sets are not degenerate). Interestingly, these optimal sets depend on $\beta$. For $\beta=1$ and $n=2$, the optimal set is the one that fragments the network the most (as in the study above with $S$ as the severity measure). For $\beta=0.066$, on the other hand, the optimal set for $n=2$ is the nodes whose removal would decrease the number of links most (even though after deleting them, the network is still connected). We can understand this since a sparser network gives fewer chances for contagion to occur, and thus a higher chance of the outbreak dying out early (which then decreases the average outbreak size).

\section{Conclusions}

We have investigated the problem of ranking influential spreaders on networks, using disease spreading as a working example. Instead of finding a quick heuristic to rank vertices in order of how influential they are, we use exhaustive search of every set of $n$ nodes to delete to find the sets that decrease the severity of the spreading the most. We find that the optimal set of $n$ nodes to delete does not in general correspond to the optimal set of $n-1$ nodes to delete augmented by just one extra node. Indeed, in practice this ill-definedness of the ranking problem can be rather severe (up to half of the optimal sets would not carry over to the next value of $n$ for the empirical network of Fig.~\ref{fig:hiv}). Our study does not necessarily disqualify papers proposing rankings of influential nodes. Indeed, for heterogeneous networks---which most real-world networks are---picking the top $n$ nodes of a ranking is probably rather close to the optimal. On the other hand, to properly evaluate ranking methods, one needs to take this issue into consideration.

The obviously most interesting question we leave open is how these results extend to larger networks (the exhaustive search used here limits us to very small networks and values of $n$). However, nothing suggests that the observed effect would vanish in larger networks. Disease spreading in metapopulations essentially follows the same model, and in such settings the ill-definedness and degeneracy of optimal sets could be relevant with very small networks (networks of farms connected by transport of livestock being one example~\cite{colizza}).

\bibliographystyle{abbrv}
\bibliography{illdef}

\end{document}